# Nonpotential Solution of the Electron Problem

## Alexander Ivanchin

### Tomsk Scientific Center

Modern physics makes wide use of the equation $\text{div}(\mathbf{E}) = \delta(\mathbf{r})$ for which only a potential solution is sought. The probability that this equation has a nonpotential solution is omitted from consideration automatically without any explanation. In this paper, using the electron problem as an example, an exact nonpotential solution linearly independent of the classical one is found. A solution with minimum potential energy is chosen as a physically feasible solution. The force acting on an electron with a nonpotential field in a homogeneous electric field and the interaction energy of two electrons with a nonpotential field is determined. The interaction energy depends on three angles responsible for mutual orientation of the electrons. Taking into account a nonpotential solution explains why there exist two types of neutrons: a neutron and an antineutron, why annihilation of the antineutron – proton is possible whereas that of the proton – neutron is impossible, why neutrons behave similar to protons. All the above properties result from the nonpotential character of an electric field.

## 1 Introduction

In theoretical physics, not only the solution of differential equations is important, but the number of solutions as well. In quantum mechanics, as an example, boundary problems normally have several solutions and all these solutions are used to describe the behavior of an object. Unreasonably discarding any of the solutions excludes one or another effect from consideration, and the description becomes inadequate and contrary to fact. If a boundary problem has several solutions and only one of them is chosen, we should justify why we do so. For practically important boundary problems, the question of solution nonuniqueness as such is often not posed. It is thought that any obtained solution is automatically unique in accordance with the Cauchy−Kovalevskaya theorem [1]; however, the theorem considers uniqueness for a boundary problem only at a certain point and not over the entire domain. Proofs of uniqueness are provided for particular differential equations, e.g., two-dimensional potential boundary problems. However, conditions of the uniqueness theorem, e.g. boundedness of a solution or problem domain, may contradict actual physical phenomena. In this work, using the field of a concentrated charge (electron, proton, etc.) as an example, it is shown that the well-known equation of mathematical physics has the second, earlier unknown solution. Taking into account this solution radically alters the interpretation of some physical processes. Surely, uniqueness of the solutions of differential equations is studied by mathematicians [2-8], but their interest (and this is quite natural) is in mathematical rather than in physical problems. The solution uniqueness of the electron problem has not been studied so far.

## 2 Nonpotential solution

The electrostatic field of a point source obeys the equation:

$$\text{div}(\mathbf{E}) = -\frac{e}{4\pi\varepsilon}\delta(\mathbf{r}), \qquad \mathbf{E} \to 0, \ \mathbf{r} \to \infty \tag{2.1}$$

where $\mathbf{E}$ is the electrostatic field, $e$ is the electron charge, $\varepsilon$ is the dielectric constant, $\mathbf{r}$ is the radius-vector, and $\delta$ is the delta (Dirac) function. It is assumed that $\mathbf{E}$ is a potential vector with $\Phi(r)$:

$$\mathbf{E} = \text{grad}(\Phi) \tag{2.2}$$

Equations (2.1) and (2.2) give the Poisson equation ($\Delta$ is Laplacian):

$$\Delta(\Phi) = -\frac{e}{4\pi\varepsilon}\delta(r) \tag{2.3}$$

The solution of (2.3) is the following:

$$\Phi = \frac{e}{4\pi\varepsilon\, r} \quad . \tag{2.4}$$

In spherical and Cartesian coordinates, the field vector **E** is written as:

$$\mathbf{E} = \frac{e}{4\pi\varepsilon\, r^3}\mathbf{r}, \quad E_r = \frac{e}{4\pi\varepsilon\, r^2}, \quad E_\theta = E_\varphi = 0 \quad . \tag{2.5}$$

The spherical and Cartesian coordinates are related as:

$$x = r \cdot \cos(\varphi)\cdot \sin(\theta), \quad y = r \cdot \sin(\varphi) \cdot \sin(\theta), \quad z = r \cdot \cos(\theta)$$

The assumption about field potentiality (2.1) has no physical grounds and limits the class of functions that can be a solution of (2.1). Hence, one needed to prove that (2.5) is a unique solution of equation (2.1), but this was not done. If there is any other solution of (2.1), it is necessary to justify the choice of (2.1) as a physically feasible solution.

Let us find a different solution of (2.1) than potential solution (2.1) by the procedure described in [10,11]. The auxiliary vector has the form:

$$\mathbf{U}^+ = -\left\{\frac{x}{r^3}, \frac{y}{r^3}, 0\right\}.$$

For the vector:

$$\mathrm{div}(\mathbf{U}^+) = \frac{1 - 3\cdot\cos^2\theta}{r^3}.$$

The vector $\mathbf{U}^+$ is neither purely potential nor purely vortex because its divergence and rotor are not identically equal to zero. If we manage to find a solution for the equation:

$$\Delta\Phi^* = \frac{1}{r^2}\cdot\frac{\partial}{\partial r}\left(r^2\cdot\frac{\partial\Phi^*}{\partial r}\right) + \frac{1}{r^2\sin(\theta)}\cdot\frac{\partial}{\partial\theta}\left(\sin(\theta)\frac{\partial\Phi^*}{\partial\theta}\right) + $$
$$+ \frac{1}{r^2\sin^2(\theta)}\frac{\partial^2\Phi^*}{\partial\varphi^2} = \frac{1 - 3\cos^2(\theta)}{r^3} \tag{2.6}$$

the vector

$$\mathbf{U} = C\left[\mathbf{U}^+ - \mathrm{grad}(\Phi^*)\right] \quad , \tag{2.7}$$

where $C$ is an arbitrary constant, is the solution of equation (2.1). We can put without sacrifice of generality:

$$C = -e/2\pi\varepsilon \quad ,$$

and the arbitrariness of $C$ is taken into account by free constants in (3.1). Then, (2.7) is written as:

$$\mathbf{U} = \frac{e}{2\pi\varepsilon}\left[\mathrm{grad}(\Phi^*) - \mathbf{U}^+\right] \quad . \tag{2.8}$$

Equation (2.6) is solved by the variable separation method:

$$\Phi^* = R(r)\cdot\Phi_1(\theta)\cdot\Phi_2(\varphi) \quad . \tag{2.9}$$

The function $\Phi^*$, as well as $\Phi$ in (2.4), must depend on $r$, and hence we put $R(r) \sim 1/r$ and the first term in (2.1) vanishes. The right side of (2.1) is invariant with $\varphi$. Let $\Phi_2 = \mathrm{const}$. Hence, the third term in (2.1) vanishes and (2.1) takes the form:

$$\frac{d}{d\theta}\left(\sin(\theta)\frac{d\Phi_1}{d\theta}\right) = (1 - 3\cos^2(\theta))\cdot\sin(\theta) \quad .$$

Its integration with respect to $\theta$ gives:

$$\frac{d\Phi_1}{d\theta} = -\sin(\theta)\cdot\cos(\theta) + \frac{A}{\sin(\theta)},$$

where $A$ is an arbitrary constant. Further integration yields:

$$\Phi_1 = \frac{1}{2}\cos^2(\theta) + A\cdot\ln\left(\tan\frac{\theta}{2}\right) \quad . \tag{2.10}$$

Although $\Phi_1$ is an unbounded function at $\theta = 0$, the field vector $\mathrm{grad}\,\Phi_1$ produced by the function is bounded at this point. It can be shown (which is done at the end of Section 2) that one of the minimum energy condition $A = 0$, and hence the second term in (2.10) is excluded from consideration. As a result, we have $\Phi_1 = (\cos^2\theta)/2$ or in Cartesian coordinates:

$$\Phi_1 = \frac{z^2}{2(x^2 + y^2 + z^2)}.$$

From the above relation and from (2.9) follows:

$$\Phi^* = \frac{z^2}{2(x^2+y^2+z^2)^{3/2}} = \frac{z^2}{2r^3},$$

and from (2.8) we obtain:

$$\mathbf{U} = \frac{e}{4\pi\varepsilon}\frac{z^2 - 2x^2 - 2y^2}{(x^2+y^2+z^2)^{5/2}}\mathbf{r}, \qquad U_r = \frac{e}{4\pi\varepsilon}\frac{2-3\cos^2\theta}{r^2} \quad U_\theta = U_\varphi = 0 \quad . \tag{2.11}$$

Substitution of (2.11) in (2.1) proves that (2.11) is a solution of (2.1). The vector $\mathbf{U}$ is nonpotential, because $\mathrm{rot}\,\mathbf{U} \not\equiv 0$. The vector $\mathbf{U}$ is axisymmetric about the $z$ axis, since $U_r$ does not depend on $\varphi$.

The flux of the vector $\mathbf{U}$ through a closed sphere centered at the origin of the coordinates:

$$\oint_S U_r\, dS = -\frac{e}{4\pi\varepsilon}\int_0^{2\pi}d\varphi\int_0^\pi U_r\, r^2 \sin\theta\, d\theta = -\frac{e}{\varepsilon} \quad . \tag{2.12}$$

Flux (2.12) is constant and is independent of the sphere radius. This suggests that

$$\mathrm{div}\,U = -\frac{e}{4\pi\varepsilon}\delta(r) \quad ,$$

as is for $\mathbf{E}$ in (2.1). From (2.1) and (2.11) it follows that the vectors $\mathbf{E}$ and $\mathbf{U}$ are parallel.

## 3 Energy

The vectors $\mathbf{U}$ and $\mathbf{E}$ satisfy boundary problem (2.1). A physically feasible solution should be chosen using the minimum energy principle, i.e., a system is in equilibrium where it has minimum energy. Consider the linear combination of the vectors:

$$\mathbf{T} = \lambda_1\mathbf{E} + \lambda_2\mathbf{U} \quad , \tag{3.1}$$

where $\lambda_1, \lambda_2$ are arbitrary constants. The choice of the constants may not affect the charge and the charge must remain constant and equal to the electron charge. This condition is met for:

$$\lambda_2 = 1 - \lambda_1 \tag{3.2}$$

The potential energy of electrostatic field (3.1) is specified by the relation [12]:

$$W = \frac{\varepsilon}{2}\int_V (\lambda_1\mathbf{E}_* + \lambda_2\mathbf{U}_*)^2\, dV \quad . \tag{3.3}$$

Using (2.1) and (2.11), we obtain:

$$W = \frac{e^2}{32\pi^2\varepsilon}\int_V (\lambda_1 E_r + \lambda_2 U_r)^2\, dV \quad , \tag{3.4}$$

The integration is performed over the entire space. Relation (3.4) is broken down into three terms. The first term:

$$W_E = \frac{\lambda_1^2}{32\pi^2} \frac{e^2}{\varepsilon} \int_V E_r^2 dV = \frac{\lambda_1^2}{32\pi^2} \frac{e^2}{\varepsilon} \int_\rho^\infty \frac{1}{r^2} dr \int_0^{2\pi} d\varphi \int_0^\pi \sin\theta\, d\theta = \frac{\lambda_1^2}{8\pi\rho} \frac{e^2}{\varepsilon} \qquad (3.5)$$

is the self-energy for the classical potential electron component. The second term:

$$W_U = \frac{\lambda_2^2}{32\pi^2} \frac{e^2}{\varepsilon} \int_V U_r^2 dV = \frac{\lambda_2^2}{32\pi^2} \frac{e^2}{\varepsilon} \int_\rho^\infty \frac{dr}{r^2} \int_0^{2\pi} d\varphi \int_0^\pi [2-3\cos^2\theta]^2 \sin\theta\, d\theta = \frac{9}{40\pi} \frac{\lambda_2^2}{\rho} \frac{e^2}{\varepsilon} \qquad (3.6)$$

is the self-energy for the nonpotential component. The third term:

$$W_B = \frac{\lambda_1 \lambda_2}{16\pi^2} \frac{e^2}{\varepsilon} \int_V (E_r U_r) dV = \frac{\lambda_1 \lambda_2}{16\pi^2} \frac{e^2}{\varepsilon} \int_\rho^\infty dr \int_0^{2\pi} d\varphi \int_0^\pi \frac{2-3\cos^2\theta}{r^2} \sin\theta\, d\theta = \frac{\lambda_1 \lambda_2}{8\pi\rho} \frac{e^2}{\varepsilon} \qquad (3.7)$$

is the interaction energy for the potential and nonpotential components. In formulae (3.5) - (3.7), the cutoff radius $\rho$ is introduced because $U_r \to \infty$   $E_r \to \infty$ at $r \to 0$. In physical terms, $\rho$ is the distance from the electron center beginning at which we can no longer speak about the electrostatic self-field of the electron.

Summation of (3.5), (3.6), and (3.7) in view of (3.2) gives the total energy:

$$W = \frac{e^2}{40\pi\rho\varepsilon} \left(9\lambda_1^2 - 13\lambda_1 + 9\right) \quad . \qquad (3.8)$$

For $\lambda_1 = 1$ and $\lambda_1 = 0$, relation (3.8) gives the energy of classical field (2.1) and that of nonpotential field (2.11), respectively. The minimum energy is reached where

$$\lambda_1 = \frac{13}{18}$$

and it is equal to

$$W_m = \frac{31}{288} \frac{e^2}{\rho\pi\varepsilon} \qquad (3.9)$$

The classical to minimum energy ratio:

$$\frac{W_E}{W_m} = \frac{36}{31}$$

That is taking into account the nonpotential component decreases the electron self-energy by near 14% compared to the classical case.
The field of minimum energy $\tilde{\mathbf{E}}$ is written as:

$$\tilde{\mathbf{E}} = \frac{e}{24\pi\varepsilon} \frac{x^2 + y^2 + 6z^2}{(x^2+y^2+z^2)^{5/2}} \mathbf{r} \qquad \tilde{E}_r = \frac{e}{24\pi\varepsilon r^2}(6 - 5\sin^2\theta), \quad \tilde{E}_\theta = \tilde{E}_\varphi = 0 \quad . \qquad (3.10)$$

Figure 1 shows the angular dependence of $\tilde{\mathbf{E}}$. The angle $\theta$ is plotted on the $z$ symmetry axis. The length of arrows is proportional to $\tilde{\mathbf{E}}$.

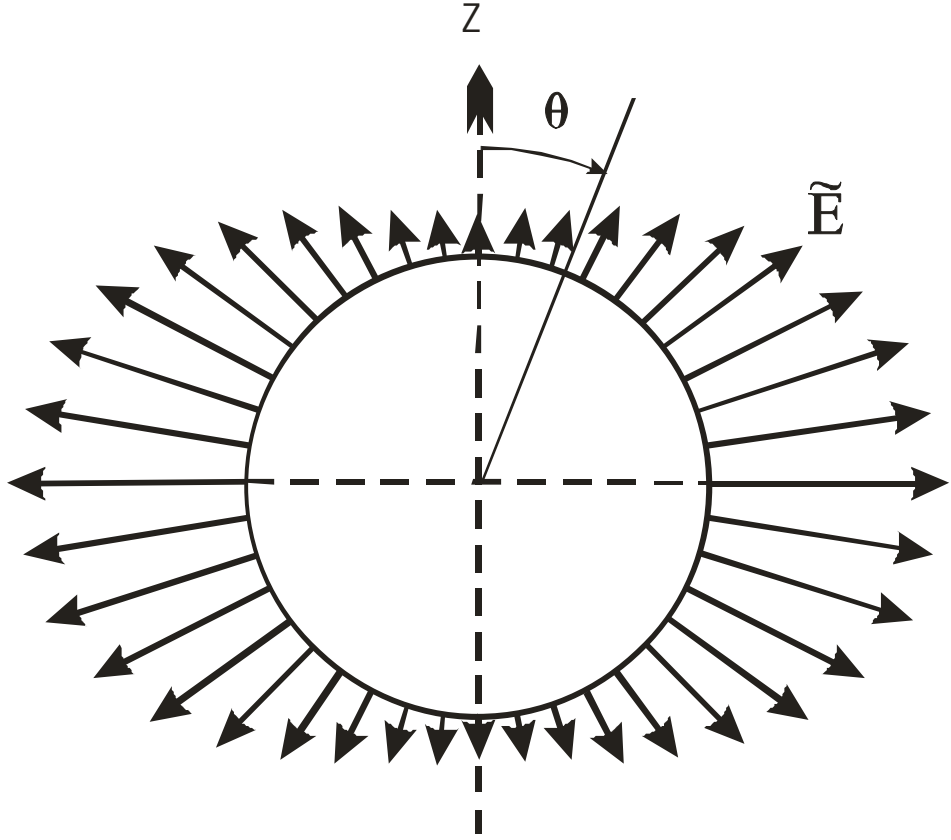

Figure 1

With two components in (3.3), the electron self-energy at the equilibrium state can have arbitrary values on retention of the charge. This is attainable by properly choosing $\lambda_1$ in (3.8). The classical solution does not provide the above possibility.

Let explain why we put $A = 0$ in (2.1). In spherical coordinates, the gradient of the second term in the right side of (2.1):

$$\mathrm{grad}\left(A \cdot \ln\left(\tan\frac{\theta}{2}\right)\right) = \frac{A}{r^2}\left\{-\ln\tan\frac{\theta}{2},\ \frac{1}{\sin\theta},\ 0\right\} \quad . \tag{3.11}$$

Then, the interaction energy for the classical component is written as:

$$-\frac{Ae}{4\pi\varepsilon}\int_0^\infty \frac{1}{r^2}dr \int_0^\pi \ln\tan\frac{\theta}{2}\sin\theta d\theta \int_0^{2\pi} d\varphi = 0$$

Interaction energy (3.11) for nonpotential component (2.11) is zero, too. Hence, minimum energy (2.3) is reached with $A = 0$.

## 4  Electron with arbitrary orientation

An arbitrary rotation can be realized by three rotations about the coordinate axes. First, a rotation about the $z$ axis through an angle $\alpha$ is made. In this rotation, the $x$ and $y$ axes take the positions $x'$ and $y'$. The second rotation is made about the $y'$ axis through an angle $\beta$ and the $x'$ axis assumes the position $x''$. The third rotation is made about the $x''$ axis through an angle $\gamma$.

For the charge at the origin of the coordinates, the initial orientation is chosen such that the symmetry axis coincides with the $z$ axis. In this case, rotation about the $z$ axis through any angle $\alpha$ changes nothing. Therefore, it can be assumed that $\alpha = 0$. The matrices for the rotations about the $y$ and $x$ axes are written as [13]:

$$\mathbf{A}_y = \begin{vmatrix} \cos\beta & 0 & \sin\beta \\ 0 & 1 & 0 \\ -\sin\beta & 0 & \cos\beta \end{vmatrix} \quad \text{and} \quad \mathbf{A}_x = \begin{vmatrix} 1 & 0 & 0 \\ 0 & \cos\gamma & -\sin\gamma \\ 0 & \sin\gamma & \cos\gamma \end{vmatrix}.$$

The vectors are transformed using the matrices $\mathbf{A} = \mathbf{A}_x \mathbf{A}_y$ and the coordinates using $\mathbf{A}' = \mathbf{A}'_y \mathbf{A}'_x$. Transformation of (3.10) gives:

$$\tilde{\mathbf{E}} = \frac{e}{24\pi\varepsilon} \frac{q}{r^5} \mathbf{r} \quad , \tag{4.1}$$

where

$$q = 10xz\cos\beta\cos\gamma\sin\beta - 10xy\cos\beta\sin\beta\sin\gamma - 10yz\cos\gamma\cos^2\beta\sin\gamma +$$
$$+ 5y^2\cos^2\beta\sin^2\gamma - 5z^2(\cos^2\gamma\sin^2\beta + \sin^2\gamma) + 5x^2\sin^2\beta + x^2 + y^2 + 6z^2$$

Because of the stress field symmetry, the angles are variable as follows:
$$0 \le \beta \prec \pi \quad 0 \le \gamma \prec \pi \quad . \tag{4.2}$$

## 5 Electron in a homogeneous field

The interaction energy of an arbitrarily oriented electron and a homogeneous external field $\hat{\mathbf{E}} = \{0, 0, \hat{E}\}$ is written as:

$$W_i = \varepsilon \int_V (\hat{\mathbf{E}} \cdot \tilde{\mathbf{E}}) \, dV = \frac{e\hat{E}}{24\pi} \int_V \frac{qx}{r^5} dV$$

The integration with the use of cylindrical coordinates $x = \rho\cos\varphi \quad y = \rho\sin\varphi \quad z = z$ gives the force:

$$f_z = \frac{\partial W_i}{\partial z} = \frac{e\hat{E}}{24\pi} \int_0^{2\pi} d\varphi \int_0^\infty \frac{q z \rho}{(z^2 + \rho^2)^{5/2}} d\rho = \frac{2}{9} eE \quad . \tag{5.1}$$

The force differs from the classical case
$$f_z = eE \tag{5.2}$$
by the factor 2/9. If denote
$$\bar{e} = \frac{2}{9} e \tag{5.3}$$
then (5.1) has the same form as (5.2)
$$f_z = \bar{e} E \tag{5.4}$$

## 6 Interaction of two arbitrarily oriented generalized electrons

The replacement (shift) $x \to x - a$ in (4.1) gives the field for an arbitrary rotated charge at the point $(a, 0, 0)$:

$$\mathbf{E}_+ = \frac{e}{24\pi\varepsilon} \frac{q_+}{r_+^5} \mathbf{r}_+ \tag{6.1}$$

where

$$q_+ = 10(x-a)z\cos\beta\cos\gamma\sin\beta - 10(x-a)y\cos\beta\sin\beta\sin\gamma -$$
$$- 10yz\cos\gamma\cos^2\beta\sin\gamma + 5y^2\cos^2\beta\sin^2\gamma - 5z^2\left(\sin^2\beta\cos^2\gamma + \sin^2\gamma\right) +$$
$$+ 5(x-a)^2\sin^2\beta + (x-a)^2 + y^2 + 6z^2$$
$$r_+ = \sqrt{(x-a)^2 + y^2 + z^2}$$
$$\mathbf{r}_+ = \{x-a,\ y,\ z\}$$

The replacement $x \to x + a$, $\beta \to \beta_-$ in (4.1) and the assumption that $\gamma = 0$ gives the field of a charge located at the point $(-a,0,0)$ and rotated about the $y$ axis in the $xz$ plane through an arbitrary angle $\beta_-$:

$$\mathbf{E}_- = \frac{e}{24\pi\varepsilon}\frac{q_-}{r_-^5}\mathbf{r}_-,$$

where
$$q_- = 10(x+a)z\cos\beta_-\sin\beta_- + 5\sin^2\beta_-(x+a)^2 + (x+a)^2 + y^2 + 6z^2$$
$$r_- = \sqrt{(x+a)^2 + y^2 + z^2}$$
$$\mathbf{r}_- = \{x+a, y, z\}$$

The orientation of the coordinate system can always be chosen such that the symmetry axis of the charge at the point $(-a,0,0)$ lies in the $xz$ plane and we can thus put $\gamma = 0$ without sacrifice of generality.

The interaction energy of the two nonpotential charges in the Cartesian coordinate system is written as:

$$W_i = \frac{e^2}{(24)^2 \pi^2 \varepsilon}\int_V (\mathbf{E}_- \cdot \mathbf{E}_+)dV \tag{6.2}$$

The integration is performed over the entire space.

Let us proceed to bipolar coordinates with poles at the points $(-a,0,0)$ and $(a,0,0)$ according to the formulae [13]:

$$x = a\frac{\sinh\tau}{\cosh\tau - \cos\sigma} \quad y = a\frac{\sin\sigma\cos\varphi}{\cosh\tau - \cos\sigma} \quad z = a\frac{\sin\sigma\sin\varphi}{\cosh\tau - \cos\sigma}$$

The Jacobean for the coordinate transformation:
$$Y = a^3 \frac{\sin\sigma}{(\cosh\tau - \cos\sigma)^3}$$

$$r_+ = \sqrt{(x-a)^2 + y^2 + z^2} = a\sqrt{2\frac{\cosh\tau - \sinh\tau}{\cosh\tau - \cos\sigma}}$$

$$r_- = \sqrt{(x+a)^2 + y^2 + z^2} = a\sqrt{2\frac{\cosh\tau + \sinh\tau}{\cosh\tau - \cos\sigma}}$$

$$r_+ r_- = \frac{2a^2}{\cosh\tau - \cos\sigma}$$

In the bipolar coordinates, the integration is first performed with respect to $\varphi$, then with respect to the variable $\tau$, and finally with respect to $\sigma$. We obtain:

$$W_i = W_A + W_C, \tag{6.3}$$

here

$$W_A = \left(\frac{5}{72}\right)^2 \frac{e^2}{\pi \varepsilon a}\left[\cos\gamma \sin 2\beta \sin 2\beta_- + \cos^2\beta \cos^2\beta_- \cos 2\gamma\right] =$$
$$= \left(\frac{5}{16}\right)^2 \frac{\bar{e}^2}{\pi \varepsilon a}\left[\cos\gamma \sin 2\beta \sin 2\beta_- + \cos^2\beta \cos^2\beta_- \cos 2\gamma\right]$$
(6.4)

$$\max\left[\cos\gamma \sin 2\beta \sin 2\beta_- + \cos^2\beta \cos^2\beta_- \cos 2\gamma\right] = 4/3 \; ,$$
when $\beta = \beta_- = 35{,}26°$, $\gamma = 0$
(6.5)

and
$$\min\left[\cos\gamma \sin 2\beta \sin 2\beta_- + \cos^2\beta \cos^2\beta_- \cos 2\gamma\right] = -4/5 \; ,$$
when $\beta = -55{,}77°$, $\beta_- = 55{,}77°$, $\gamma = 0$
(6.6)

$$W_C = 2\left(\frac{5}{72}\right)^2 \frac{e^2}{\varepsilon \pi a} = 2\left(\frac{5}{16}\right)^2 \frac{\bar{e}^2}{\varepsilon \pi a}$$
(6.7)

$a$ is a half of distance between electrons. For the same charges $W_C > 0$ and they repulse. For the charges of opposite signs $W_C < 0$ and they attract. The summand $W_A$ due to orientation of electrons may be positive or negative.

## 7  Comparison of classical and generalized electrons

If the measurement of electron charge is run in homogeneous field it is impossible to conclude we have deal with classical or generalized electron as (5.2) and (5.4) have the same form. To draw the conclusion about electron field structure the angle distribution of electric field intensity should be defined.

Interaction energy of classic electrons is
$$W = \frac{1}{8}\frac{e^2}{\pi \varepsilon a}$$
(7.1)

Interaction energy of generalized electrons (6.3) consists of two summands. The numerical coefficient before
$$\frac{e^2}{\pi \varepsilon a}$$

in (6.7) more then 10 times less then for (7.1). When the substitution $e \to \bar{e}$ is produced values of these coefficients become approximately equal. Complete equality is impossible as (6.7) depend on orientation of electrons as opposite to (7.1) is independent.

The orientation dependence creates maybe indeterminacy principle due to next two causes. First, electron orientation is not detected in experiments. Second, the spontaneous changing of orientation is possible when charge particles are collided with.

There is a free parameter in (3.8) and consequently the electron self-energy may differ under condition of charge conservation. The equilibrium state has minimum of energy if not the particle has exciting state. The exciting state of classic electron is used but modern physics fail to explain what it is. For generalized electron such problem doesn't exist.

## 8  Neutron

The neutron charge is equal to zero, then from (3.1) we obtain
$\lambda_2 = -\lambda_1 = \pm C$.
Here $C$ is the arbitrary constant. It may be taken equal to 1. There are two possible solutions
$\lambda_1 = 1 \quad \lambda_2 = -1$

and
$$\lambda_1 = -1 \quad \lambda_2 = 1 \,. \tag{8.1}$$
This agrees well with the known fact that there are two types of neutrons: a neutron and an antineutron. The electric field of neutrons and antineutrons is written as
$$\mathbf{N} = \pm(\mathbf{U} - \mathbf{E}) = \pm \frac{3e(x^2 + y^2)}{4\pi\varepsilon\, r^5}\mathbf{r} = \pm \frac{e}{4\pi\varepsilon\, r^2}\left(1 - 3\cos^2\theta \,,\; 0\,,\; 0\,,\right) \tag{8.2}$$

The field vector $\mathbf{N}$ has only one radial component. Depending on the angle $\theta$, it is directed either to the neutron center (let us call that region convergent) or from the neutron center (let us refer to that region as divergent).

From the solution of the equation
$$1 - 3\cos^2\theta = 0$$
we find the angle separating the regions
$$\theta_0 = \arccos\frac{1}{\sqrt{3}} \approx 54{,}7°\,.$$
Figure 2 shows schematically the solution ) for a neutron and Figure 3 for an antineutron.

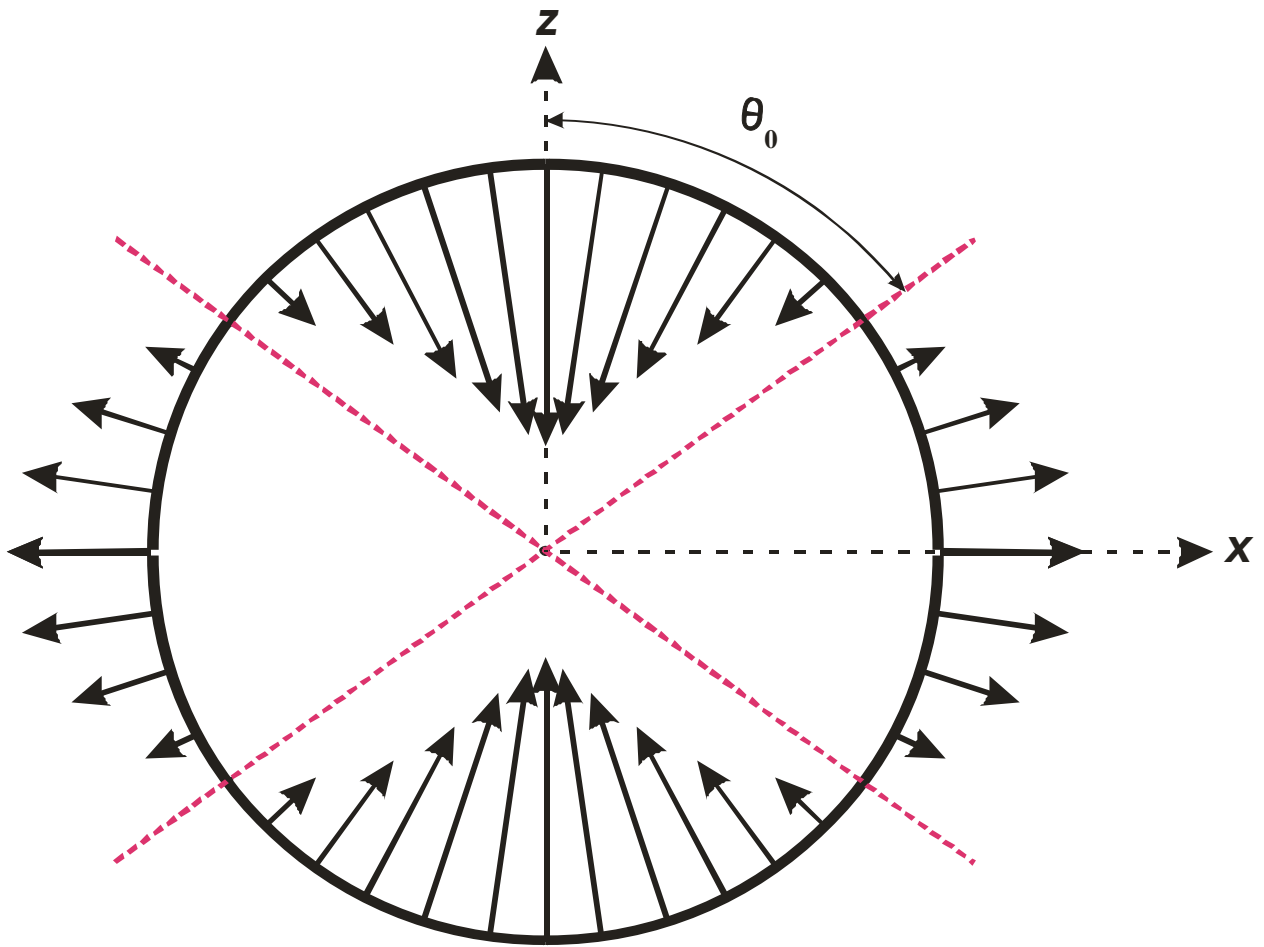

Figure 2

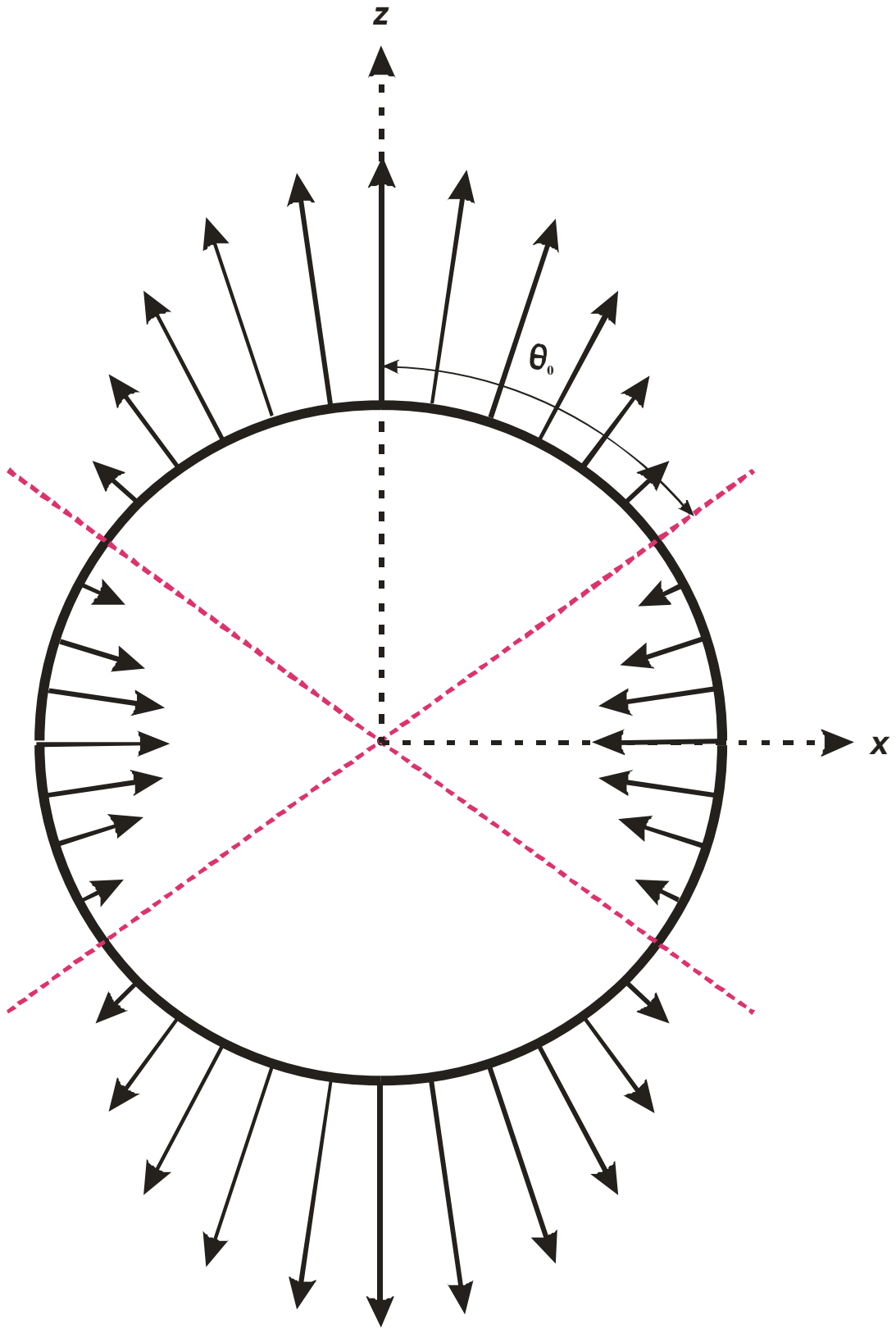

Figure 3

The angle $\theta_0$ is not multiple of $\pi/n$, here $n$ is a natural number. Therefore, it is not possible to transform a neutron into an antineutron and vice versa by turning the coordinates. These are, in fact, two different particles. The neutron self-energy is written as

$$W_N = \frac{e^2}{32\pi^2 \varepsilon} \int_\rho^\infty \frac{dr}{r^2} \int_0^\pi (1 - 3\cos^2 \theta)^2 \sin\theta\, d\theta \int_0^{2\pi} d\varphi = \frac{e^2}{5\pi\varepsilon\rho_N}. \tag{8.3}$$

The radius of cutting for a neutron is determined by its rest mass $m_N$ from the equality

$$\frac{e^2}{5\pi\varepsilon\rho_N} = m_N c^2.$$

Hence

$$\rho_N = \frac{e^2}{5\pi\varepsilon\, m_N c^2}.$$

Let us find the energy of the neutron convergent and divergent regions. From (8.3) we have

$$\frac{e^2}{32\pi^2\varepsilon}\int_\rho^\infty \frac{dr}{r^2}\int_0^{2\pi}d\varphi \int_{\pi-\theta_0}^{\theta_0}(1-3\cos^2\theta)^2 \sin\theta\, d\theta = \frac{2}{3\sqrt{3}}W_N \approx 0{,}385 W_N.$$

And the energy for $\theta \in (0,\ \theta_0)\cup(\pi-\theta_0,\ \pi)$ is $\left(1-\dfrac{2}{3\sqrt{3}}\right)W_N \approx 0{,}615 W_N$.

Thus, the neutron self-energy may be distributed in the following way: one-third in the angle $\theta \in (\theta_0\ \pi-\theta_0)$ and two-thirds in the angle for $\theta \in (0,\ \theta_0)\cup(\pi-\theta_0,\ \pi)$. The energy distribution is unsymmetric. It is this distribution which is characteristic of neutrons and antineutrons. The proton field is divergent. A greater part of the antineutron field is convergent. It is for this reason that protons and antineutrons can annihilate. A greater part of the neutron field is also divergent as that of the proton, so they cannot annihilate. The electric energy of the potential proton is given by the relation

$$W_E = \frac{e^2}{8\pi\varepsilon\,\rho_P}$$

Here $\rho_P$ is the radius of cutting for the proton (the "classical" proton radius). The electric energy of the nonpotential proton is written in the same way as that of the nonpotential electron (3.9), with the electron radius of cutting replaced by the proton radius of cutting

$$W_P = \frac{31}{288}\frac{e^2}{\rho_P\,\pi\varepsilon}. \tag{8.4}$$

## 9  Interaction of a neutron with a homogeneous electric field

Let us choose the Cartesian coordinate system so that $z$ axis is directed along the neutron axis of symmetry and $x$ axis is directed so that the coordinate of the external field vector $\tilde{E}_y = 0$. In the above coordinate system the uniform external field looks like

$$\tilde{\mathbf{E}} = \{\tilde{E}_x,\ 0,\ \tilde{E}_z\}.$$

It is obvious that such a choice is always possible. The neutron field in such coordinated is written in the form of (8.2). The interaction energy of the arbitrarily turned neutron with a uniform electric field is written as

$$W_i = \varepsilon\int_V (\mathbf{N}\cdot\tilde{\mathbf{E}})dV = -\frac{3e}{4\pi}\int_0^\infty dr\int_0^\pi \sin^3\theta\, d\theta \int_0^{2\pi}(\tilde{E}_x\cos\varphi\sin\theta + \tilde{E}_z\cos\theta)d\varphi = 0.$$

The interaction energy of a neutron and a homogeneous electric field is equal to zero. This means that the force acting on the neutron on the side of the homogeneous electric field is also equal to zero. The same equality may not be true for the nonuniform external field. That a neutron, under certain conditions, interacts with an electric field is an experimental fact [14].

## 10 The neutron - neutron interaction

The field of the arbitrarily directed neutron is written as

$$\mathbf{N} = \pm \frac{e}{24\pi\varepsilon r^3}\left(\frac{q}{r^2}-1\right)\mathbf{r} = \pm \frac{5e}{24\pi\varepsilon r^5}\{[-2yz\cos\gamma + (y^2-z^2)\sin\gamma]\cos^2\beta\sin\gamma +$$
$$+ 2x[z\cos\gamma - y\sin\gamma]\cos\beta\sin\beta + x^2\sin^2\beta + z^2\}\mathbf{r} \tag{10.1}$$

Just as in (6.1), we locate two neutrons of an arbitrary direction at the point of abscissa $x = a$ and $x = -a$. The strength of the neutron electric field at $x = a$ is written by analogy with (6.1) as

$$\mathbf{N}_+ = \frac{5e}{24\pi\varepsilon r_+^5}\{[-2yz\cos\gamma + (y^2-z^2)\sin\gamma]\cos^2\beta\sin\gamma +$$
$$+ 2(x-a)[z\cos\gamma - y\sin\gamma]\cos\beta\sin\beta + (x-a)^2\sin^2\beta + z^2\}\mathbf{r}_+ \tag{10.2}$$

For the neutron located at $x = -a$

$$\mathbf{N}_- = \frac{5e}{24\pi\varepsilon r_-^5}\{2(x+a)z\cos\beta_-\sin\beta_- + (x+a)^2\sin^2\beta_- + z^2\}\mathbf{r}_-$$

The energy of the neutron – neutron interaction is founded by analogy with (6.2) as
$W_{NN} = W_A$,
here $W_A$ is (6.4).

If one calculates the neutron-antineutron interaction energy, then one should take the sign minus in $W_{NN}$.

## 11 Proton-neutron interaction

The interaction energy of the neutron located at the point $+a$ on abscissa and the proton (6.1) located an point $-a$ is written as

$$W_{PN} = W_A + \frac{8}{5}W_C$$

The energy of the proton-neutron interaction consists of $W_A$ energy and $W_C$ energy, with $W_C$ energy here 1,6 times larger than that for the proton-proton interaction (8.4). That is, there is a repulsive component in the force of the neutron-proton interaction which agrees with the known experimental fact that protons and neutrons are approximately alike in their interaction. Even for this reason, they are combined in the same group called nucleons. Protons interact via an electric field. According to the potential theory, a neutron does not possess such a field and cannot interact with charged particles. From this point of view, the similarity of protons and neutrons is not clear. If one takes into account the nonpotential part of an electric field, then the similarity in the behaviour of neutrons and protons becomes natural.

## 12 Conclusion

The solution nonuniqueness for equation (2.1) demonstrated in the paper has not been studied so far. In this context, several questions arise: how many solutions for the boundary problem exist; what the conditions for alterative solutions are; what conditions other than minimum energy, if any, provide uniqueness. These questions are relevant not only to the electron theory, since equations of type (2.1) occur in different physical problems, such as those of mechanics of liquid and gas, theory of elasticity, gravity, etc. In mathematical terms, the question of uniqueness is undeniably important and the fact that solution (2.1) was historically the first is likely a chance. If solution (2.1) rather than, e.g. (2.11), was found and this solution was assumed unique, one would consider it unique to describe the interaction of charged particles.

As it turned out, the minimum potential energy is not achieved with the potential solution. It is achieved at some linear combination of potential and nonpotential solutions. The nonpotential solution possesses axial symmetry unlike the potential one, which spherically symmetric. The absence of minimum in the known solution is a challenging problem in theoretical physics. This means that the principle of least action does not work in the potential solution which is true not only for some exotic particles but also in the case of protons, electrons, etc. There can be three solutions of the above problem:

To abandon the principle of least action as the wrong one.

There exists some unknown physical law in addition to the principle of least action, which restricts the choice of solutions only to potential functions. Therefore, from among all solutions one has to choose only potential ones.

To consider that the principle of least action is fulfilled and use it to determine the distribution of the electric field strength near the elementary charge taking into account nonpotential solutions as those that can be physically realized.

If the principle of least action were wrong it would be clear in experiments because it would not be fulfilled on the main particles that our world consists of. Therefore, the first variant should be rejected. For the same reason, the second variant should also be dropped. Hence the third variant is consided here.

It is impossible to understand why there exist two types of neutrons without taking into account the nonpotential component of an electric field. It is clear in the case of the proton and the antiproton: opposite charges cause attraction between them, which leads to annihilation. It is not clear why a proton can annihilate into an antineutron. In the potential approach there is no attraction force between them. Taking into account the nonpotential part of an electric field eliminates these problems making the existence of the two types of neutrons a natural thing. Between a proton and an antineutron there is a negative interaction energy that is they attract each other.

From the above calculations it follows that there is a negative interaction energy between a neutron and an electron resulting in their attraction. It corresponds to the fact that neutrons get reflected from atomic nuclei rather than from electron shells. It is not clear if the formation of the related pairs of the type of antineutron-proton and neutron-positron is possible by analogy with the hydrogen atom. There exists the attraction between them.

In the theory of the electromagnetic field ones concludes from the equality $\nabla \cdot \mathbf{B} = 0$ that there is such vector-potential that $\mathbf{B} = \nabla \times \mathbf{\Psi}$. There is a similar assumption for an electric field, which is known as the potential gage invariance. The above assumptions are correct, and such potentials really exist. It should only be noted that along with them, there also exist nonpotential solutions. They should be taken into account in a consistent theory. It can be shown that a set of the nonpotential functions satisfying the continuity equation has the power of continuum, that is such functions are not few, as many as potential functions. A complete solution of the Maxwell equations will be their linear combination. This second solution is also important. For example, it is responsible for quantum effects. Besides, it is impossible to describe the neutron properties without it, etc. The nonpotential solution is not taken into account in problems of confining plasma, which may be one of the reasons for which it is still impossible to confine it for carrying out a thermonuclear reaction.

In this stage it is not clear how one can derive quantum equations from Maxwell's electromagnetic equations using nonpotential solutions.

In the analysis performed here the magnetic component was not taken into account. From Maxwell's equation

$$\frac{\partial \mathbf{B}}{\partial t} = -\operatorname{rot} \mathbf{E} \qquad (12.1)$$

it follows that the curl of an electric field brings about a nonstationary magnetic field. For a nonpotential solution the curl is not equal identically to zero $\operatorname{rot} \mathbf{E} \not\equiv 0$.

Should one introduce into (12.1) an additional component that ensures existence of a curl electric field for a stationary process, as is the case with the second Maxwell's equation, where along with the magnetic field curl there is a term taking into account an electric current? Isn't the magnetic dipole of elementary particles this term? Really, taking a rotor from (3.10) we obtain the following in the spherical coordinate system

$$\text{rot } \mathbf{T} \sim \frac{\sin 2\theta}{r^3}\{-\sin\varphi, \quad \cos\varphi, \quad 0\}$$

The magnetic dipole field has exactly the same functional dependence.

An important question is how many solutions of the continuity equation there exist (2.1). Nobody has answered this question yet. There are only some considerations concerning this problem. As is shown in [15], an exact solution of three different problems of the elasticity theory is possible only if two solutions are taken into account: potential and nonpotential. If there were other solutions, they would have manifested somehow, so the exact solution of the problems would not have been obtained. The above fact suggests that there are only two linearly independent solutions. There are only two neutrons so. But on the other hand, two-dimensional problems of the elasticity theory were considered in [15]. For elementary particles it is necessary to solve three-dimensional problems. The first solution, which is potential, has a spherical symmetry. The second, which is nonpotential, possesses an axial symmetry. Can there be the third unsymetric solution? So far nobody has been able to prove mathematically that here exist only two or more solutions.

The results of taking into account nonpotential solutions look unusual from the viewpoint of modern concepts. Neutrons, as it turned out, possess an electric field and interact with each other. In Maxwell's electrodynamics equations one should take into account a previously unknown group of solutions. Also, there appear many other problems. However, as it has been demonstrated in this study, taking into account a nonpotential solution makes the neutron properties, which are impossible to explain in terms of the potential theory, quite clear.